\documentstyle[11pt,newpasp,twoside,epsf]{article}
\def\edcomment#1{\iffalse\marginpar{\raggedright\sl#1\/}\else\relax\fi}
\reversemarginpar

\begin{document}
\title{AGB stars in the Magellanic Clouds \& in other members of the Local Group}
 \author{Maria--Rosa L. Cioni}
\affil{European Southern Observatory, Karl--Schwarschild--Str. 2, 
D--85748 Garching bei Muenchen, Germany}

\begin{abstract}
Results obtained in the Magellanic Clouds using the latest near--IR DENIS 
survey are briefly revised. This sets the base to a similar study of AGB 
stars in other galaxies in the Local Group and in particular in 
NGC6822.
\end{abstract}

\section{Introduction}
There  are two  main advantages  of studying  asymptotic  giant branch
(AGB) stars in the galaxies of the Local Group (LG). First that we are able
using the current instrumentation to resolve their stellar content and
second that these stars can be considered to be approximately all 
at the same distance.
Recent studies  of AGB  stars in the  Magellanic Clouds allowed  us to
gain insight into the near--infrared (near--IR) stellar content of the
galaxies,   their   surface  distribution   and   the  ratio   between
carbon--rich  (C--rich) and  oxygen--rich (O--rich, M--type)  AGB  stars easily
statistically distinguished  in the colour--magnitude  diagram (CMD, 
$J-K_S$, $K_S$).
More recent near--IR observations of other galaxies in the Local Group 
have been analyzed to tell us similar informations.

\section{Results on the Magellanic Clouds}

Among the latest near--IR  instruments DENIS and  2MASS have  released a
large  amount  of data  on  AGB stars  in  the  Magellanic Clouds.  In
particular in the DENIS  catalogue towards the Magellanic Clouds (DCMC
-- Cioni et  al. 2000a)  there are  $32800$ AGB stars  in the Large 
Magellanic Cloud (LMC) and $7650$ in the Small Magellanic Cloud (SMC), 
of  which  $7570$  and  $1640$  are C--rich. Blanco  et al.  (1983) 
estimated about  $11000$ and  $2900$ C--rich
stars in  the LMC and  SMC, respectively. Accounting for  about $10\%$
misclassification among the M0--1 stars below the tip of the red giant
branch (TRGB) which are missed  by our selection criteria based on the
near--IR colour--magnitude diagram (CMD), we detect most of the AGB stars in
both  Galaxies.  Infact  we detect  about the  same C  stars  found by
Kontizas et al. (2001) in the LMC ($7750$) and those $1707$
found by Rebeirot et al. (1983) in the SMC.

\begin{figure}
\plotone{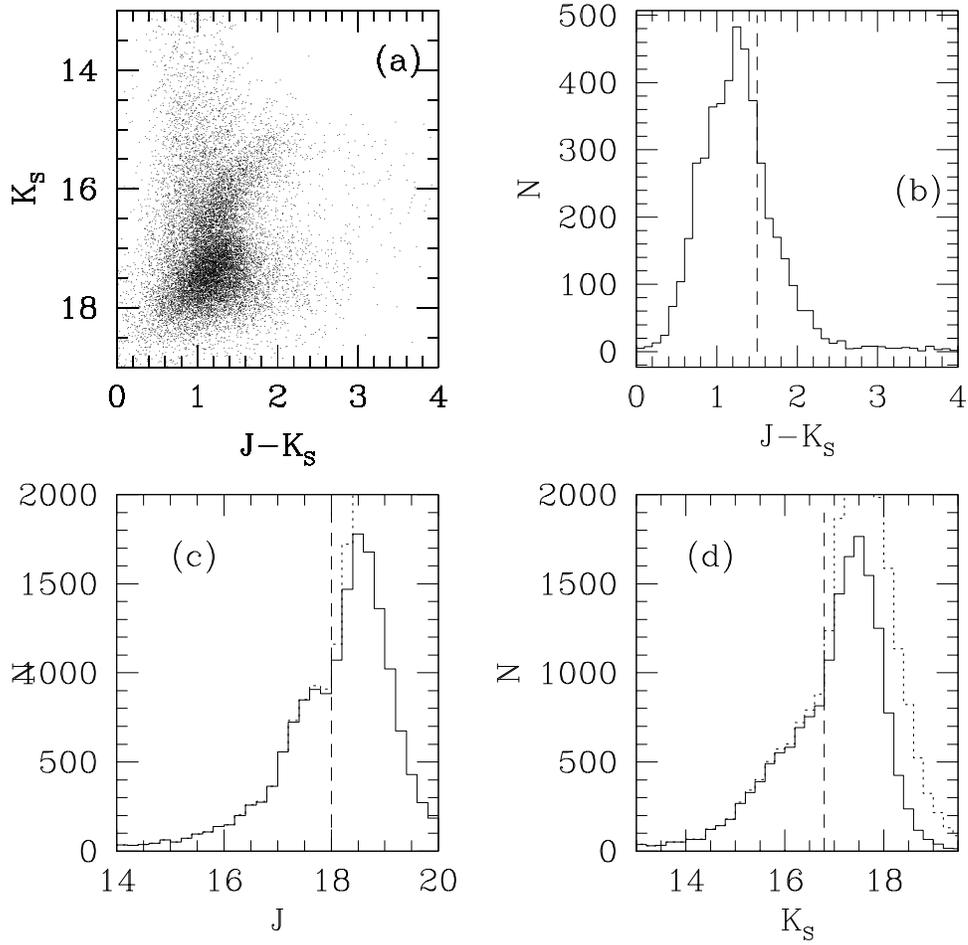}
\caption{Colour--magnitude diagram and histogram in colour and magnitudes 
for the sources observed towards NGC6822.}
\end{figure}

AGB  stars are easily  distinguished in  the CMD  ($I-J$, $I$)  as the
plume of objects above the TRGB redder of a given line that discriminates 
between
the  younger and  older populations  (Cioni et  al. 2000b).  These AGB
stars  distribute  smoothly  over  the  surface of  the  Clouds  in 
contrast with  the bar--like and  patchy distribution of  the younger
stars. In  the CMD ($J-K_S$, $K_S$)  C--rich AGB stars  occupy a 
red  branch  compared  to   the  location  of  O--rich  stars.  The
distribution of the ratio between  C--rich and O--rich (C/M ratio) AGB
stars outlines  in the LMC  a ring--like structure of increasing values
(Cioni \& Habing  2003). Because the C/M ratio  is a strong indicator 
of  metallicity the  first global  evidence that  there  is a
radial  metallicity gradient  in  the  LMC was found. 
The  ratio  is patchy  and
irregularly  distributed in the  SMC. By  fitting the  distribution of
points (Log($C/M$) versus [Fe/H]) in other galaxies of the LG 
a  metallicity  spread  of about  $0.75$  dex within  each
Cloud was derived.  
The larger  uncertainty  remains in  the  calibration and scatter 
of  this relation.

\section{Results in other Local Group galaxies}

Based on  the results obtained in  the Magellanic Clouds  I started in
collaboration with Habing a  near--infrared campaign from the La Palma
observatory to mosaic  other galaxies of the LG in $I$, $J$ and $K_S$. 
The goal is to find, study and compare the AGB stellar population.

Fig.1 shows: the ($J-K_S$, $K_S$) diagram for all the sources detected
within an  area of $20^{\prime}\times20^{\prime}$  centered on NGC6822
(a), the histogram of the $J-K_S$  colour at about $0.3$ mag above the
TRGB where the dashed line discriminates between
O--rich and  C--rich AGB stars (b), and the histograms  of $J$ (c)
and $K_S$ (d) where the dashed line indicates the approximate location
of the TRGB. In NGC6822  we detect about $1339$ C stars, approximately
$400$ more than Letarte et al.  (2002), and $2774$ M stars. The distribution
of the  whole AGB population is  shown in Fig.2 while  Fig.3 shows the
distribution of  the C/M ratio.  Using the same relation  discussed in
Sect.2 we derive that the gray  scale and the contours span a range of
1.65  dex in  [Fe/H]. 
Contrary to Nowotny et al. (2003) we do detect regions with
a different C/M ratio.
Note  that in  the $K_S$  band  the differential
reddening  is   negligible  and   that  foreground  stars   have  been
removed. The striking difference between the two figures indicates the
potential  of  the   C/M  ratio  to  study  the   chemical  history  of
galaxies. 

Similar figures,  though with less  statistics have been  obtained for
NGC147 and  NGC185. Observations of  Draco have just been  reduced and
unfortunately  the observations of  LeoA,  LeoI and  LeoII 
took place  during variable  sky conditions which  considerably affect
the quality of the resulting CMDs.

In the Southern Hemisphere the central region of a few galaxies has been 
observed using SOFI at the NTT in the near--IR wave bands by Tolstoy back 
in 1998. The CMDs of DDO210, Fornax, SagDIG and Pegasus show clear red 
giant branches reaching in Fornax the red clump. These galaxies are not 
rich in AGB stars especially of C type except Pegasus that shows a plum 
of objects with quite red $J-K_S$ colours. These relatively deep data are 
probably suitable to derive the metallicity and the age of the observed 
region (about $5^{\prime}\times5^{\prime}$) from the slope and colour of
the RGB as in Davidge (2003). New observations of these and other galaxies 
visible from the Southern Hemisphere took place at the end of July 2003 in 
collaboration also with Reijkuba.

The wealth of near--IR data on galaxies in the Local Group either than the 
Magellanic Clouds will definitely benefit from the latest theoretical results 
by Marigo et al. (2003). The authors were able to successfully model the red 
tail of C stars producing a synthetic ($K_S$, $J-K_S$) diagram in very good 
agreement with the observational data. The key ingredient is to assume an
 opacity that in cool stars varies with the chemical composition in addition 
to the assumption of a given pulsation mode (the first overtone for C stars).

\begin{figure}
\plottwo{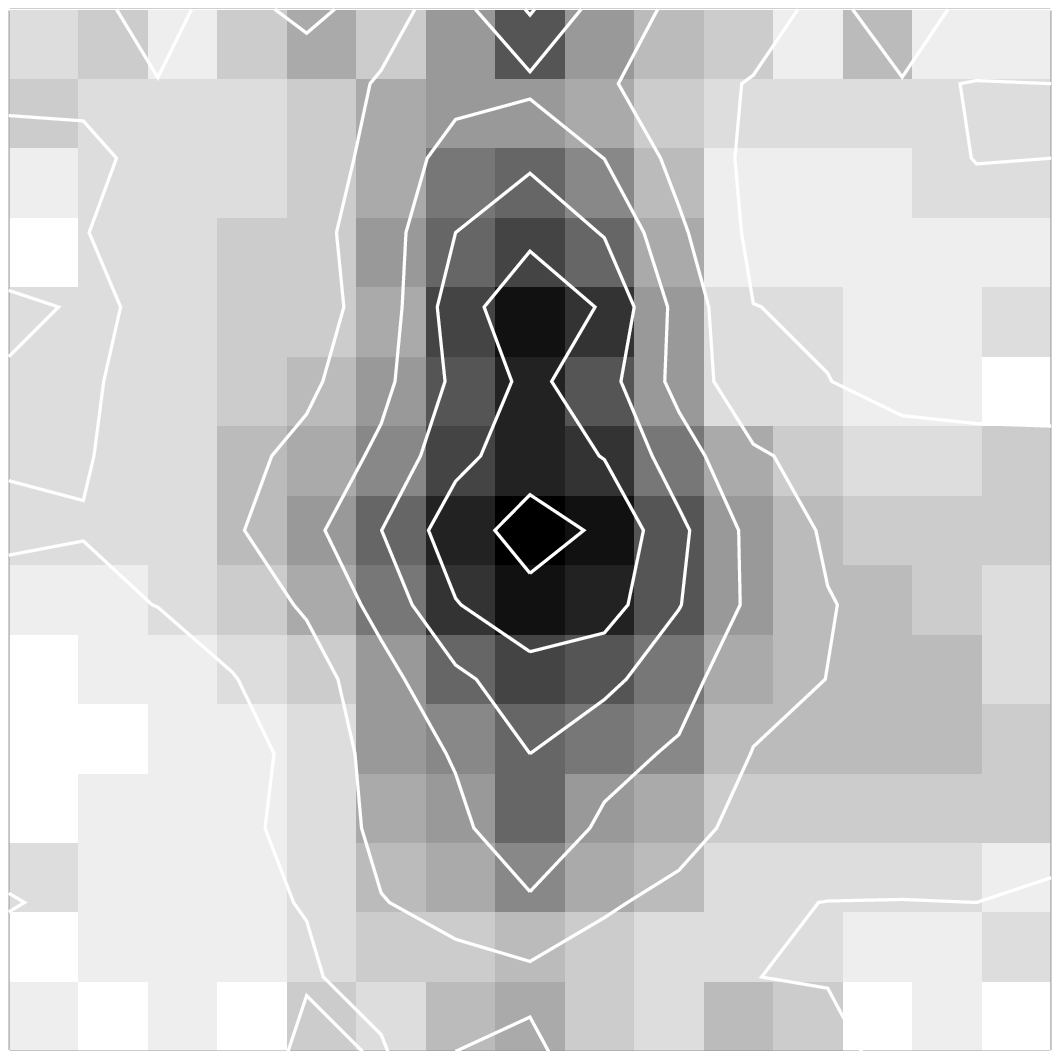}{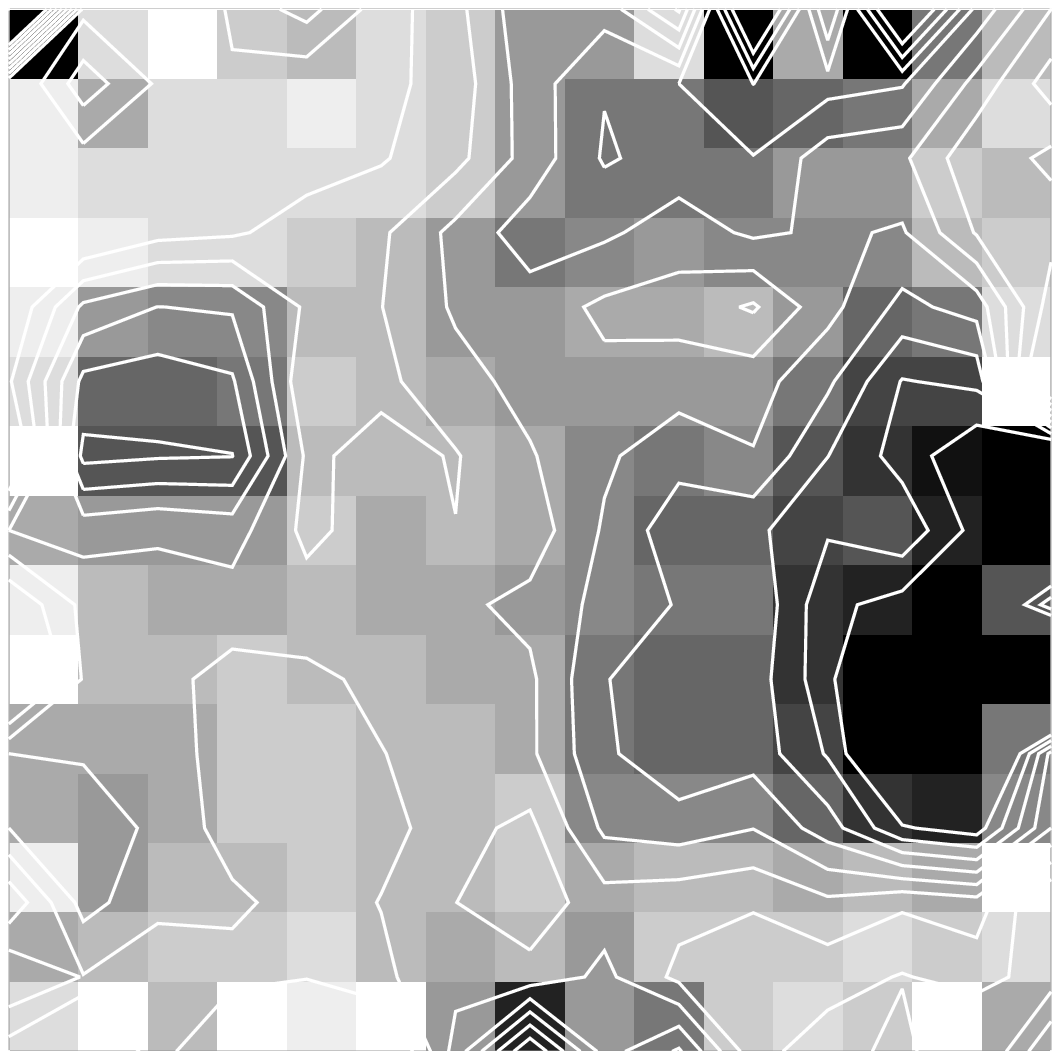}
\caption{Spatial distribution of AGB stars (left) and C/M ratio 
(right) in NGC6822.}
\end{figure}

\section{Conclusions}

Large  scale  informations on  AGB  stars  in  the Magellanic  Clouds,
NGC6822,  NGC147 and  NGC185 have  been discussed: 
their location in the near--IR CMD, their spatial distribution and that of the
C/M ratio. The latter  is an important abundance indicator, especially
for  those systems  too  far  away to  measure  abundances with  other
tracers.  In  the  near  future  we will  complete  the  reduction  of
complementary I--band measurements, try to complete the observations of
some  targets not yet fully mosaiced and to  continue a  monitoring
program  (in  the I--band)  that  will  provide us  with  an
indication about  the variability of  the target stars.  Ultimately we
count on  publishing homogeneous catalogues of AGB  variables that will
be  useful  to  the  whole  community  to perform  stellar
population studies in these relatively nearby galaxies.


\begin{references}
\reference Blanco V.M., McCarthy M.F., 1983, AJ 88, 1442
\reference Cioni M.-R.L., Habing H.J., 2003, A\&A 402, 133
\reference Cioni M.-R.L., Habing H.J., Israel F.P., 2000b, A\&A 358, L9
\reference Cioni M.-R.L., Loup C., Habing H.J., et al., 2000a, A\&A 144, 235
\reference Davidge T.J., 2003, AJ 125, 3046
\reference Kontizas E., Dapergolas A., Morgan D.H., et al., 2001, A\&A 369, 932
\reference Letarte B., Demers S., Battinelli P., et al., 2002, AJ 123, 832 
\reference Marigo P., Girardi L., Chiosi C., 2003, A\&A 403 225
\reference Nowotny W., Kerschbaum F., Olofsson H., et al., 2003, 403, 93
\reference Rebeirot E., Martin N., Prevot L., et al., 1983, A\&AS 53, 255
\end{references}
\end{document}